\documentclass{article}
\usepackage{graphicx}

\def\giorno{21 June 2007}

\def\a{\alpha}
\def\b{\beta}

\def\de{\delta}   
\def\eps{\varepsilon}
\def\la{\lambda}

\def\th{\theta}
\def\vth{\vartheta}
\def\vphi{\varphi}

\def\A{{\mathcal A}}

\def\L{{\mathcal L}}

\def\pa{\partial}
\def\d{{\rm d}}       

\def\<{\langle}
\def\>{\rangle}

\def\({\left(}
\def\){\right)}
\def\[{\left[}
\def\]{\right]}
\def\=#1{\bar #1}
\def\~#1{\widetilde #1}
\def\.#1{\dot #1}
\def\^#1{\widehat #1}
\def\"#1{\ddot #1}

\def\salta#1{{}}

\def\beq{\begin{equation}}
\def\eeq{\end{equation}}
\def\feq{\end{equation}}

\begin{document}

\title{Sine-Gordon solitons, auxiliary fields,
and singular limit of a double pendulums
chain}

\author{Mariano Cadoni\footnote{{\tt mariano.cadoni@ca.infn.it}},
Roberto De Leo\footnote{{\tt roberto.deleo@ca.infn.it}} \\
{\it  Dipartimento di Fisica, Universit\`a di Cagliari} \\
and {\it I.N.F.N., Sezione di Cagliari,} \\
{\it Cittadella Universitaria, 09042 Monserrato (Italy)} \\
 { } \\
Giuseppe Gaeta\footnote{{\tt
gaeta@mat.unimi.it}} \\
{\it Dipartimento di Matematica, Universit\`a di
Milano,} \\
{\it via Saldini 50, 20133 Milano (Italy)} }

\date{\giorno}



\maketitle

\noindent {\bf Summary.} We consider the continuum version of an
elastic chain supporting topological and non-topological degrees
of freedom; this generalizes a model for the dynamics of DNA
recently proposed and investigated by ourselves. In a certain
limit, the non-topological degrees of freedom are frozen, and the
model reduces to the sine-Gordon equations and thus supports
well-known topological soliton solutions. We consider a (singular) 
perturbative expansion around this limit and study in particular
how the non-topological field assume the role of an auxiliary
field. This provides a more general framework for the slaving of
this degree of freedom on the topological one, already observed
elsewhere in the context of the mentioned DNA model; in this
framework one expects such phenomenon to arise in a quite large
class of field-theoretical models.

\section{Introduction}

In a recent paper \cite{CGDlong} we analyzed a ``composite'' model
of DNA torsion dynamics, pretty much in the spirit of the by now
classical Peyrard-Bishop, Yakushevich and Barbi-Cocco-Peyrard
models \cite{PB,YakPLA,GRPD,YakBook,BCP,BCPR,CM,DPbook,PeyNLN};
this model amounted to a field-theoretic Lagrangian with
topological and non-topological interacting fields. We studied in
particular the solitary wave excitations (we call these {\it
solitons} for ease of language) it supports; this analysis
displayed some puzzling and somehow surprising features:
\begin{itemize}
\item[{\it (a)}] on the one hand, the speed of given soliton
solutions of a certain (relevant) type is not a free parameter but
turns out to be fixed by (the parameters of) the model; \item[{\it
(b)}] On the other hand, when performing a perturbative analysis
(exact solution of the model was not possible) the non-topological
field turns out to be completely determined -- we say then it is
{\it slaved} -- by the topological one. \end{itemize}

It appeared that these phenomena are -- rather obviously -- not
specific to the model considered there, nor to DNA dynamics, but
actually common to a much wider class of models; they could also
be of rather obvious interest in applications. We believe
therefore they are worth further investigation and clarification.

In a previous related paper \cite{CDGspeed} we have investigated
the selection of the soliton speed, i.e. point {\it (a)} above; in
the present paper we focus on point {\it (b)}, i.e. on the slaving
of one field. We identify the origin of this in the remarkable
fact that the slaved field, while being on equal footing in the
full theory, becomes an {\it auxiliary field in the perturbative
expansion}; that is, at each order in the perturbative expansion
the Lagrangian depends on the field but not on the conjugate
momentum.\footnote{From the point of view of DNA dynamics it is
interesting to remark, in this respect, that the BCP model of DNA
dynamics was recently shown \cite{GaeVen} to admit a constant of
motion in its field-theoretic limit; the presence of a constant of
motion leads to a reduction in the effective dimensionality of the
theory, similarly to what happens in the presence of an auxiliary
field.}

Auxiliary fields appear in a number of field theories (e.g.,
auxiliary fields play a crucial role in supersymmetric field
theories where they allow for off-shell closure of the algebra
\cite{bagger, nieuwenhuizen}), but it is somehow surprising to
have these in the framework of elastic chains, and even more so in
the context of DNA dynamics. Moreover, the appearance of auxiliary
fields and the related slaving mechanism are quite interesting
from the point of view of the would-be functional role of DNA
solitons, in that they would provide a robust mechanism for the
coordination of different degrees of freedom, essential if the
solitons have to play a role -- as conjectured since a long time
\cite{Englander} (see also \cite{GRPD,YakBook,DPbook}) -- in DNA
transcription.

More generally, the mechanism at work here can be described as
follows: one considers a system with two degrees of freedom, and
the (singular) limit in which one of the two fields is constrained
to zero, determining an exact solution to the full field equations
for the other field; if one considers the (singular) perturbation
expansion around this field configuration, it turns out that the
perturbation for one field are completely determined -- via
algebraic identities -- by the solutions for the dynamical
equations for the other. It appears the mechanism is -- in these
abstract terms -- quite general; we expect therefore that it can
be generalized (as the speed selection mechanism mentioned above
\cite{CDGspeed}) to a number of other physical situations.

In order to show more clearly the mechanism at work, we study an
elementary model, made of a chain of coupled double pendulums;
while the first pendulums are standard ones, free to swingle round
the circle, the second pendulums oscillations are constrained to
an angular range $|\vphi| \le \phi_0$ . We consider the continuum
limit of the chain, which yields a field-theoretic Lagrangian with
a topological and a non-topological degree of freedom
(corresponding respectively to the angular coordinates of first
and second pendulums).

\section{The double pendulums chain model}

We will consider an infinite chain of double pendulums, suspended
at points $\xi_n = n \de$ on a straight horizontal line, denoted
below as the ``axis'' of the chain.

Each of the double pendulums is made of a first massless rigid
beam of length $R$ suspended at one extremum on the line and
having a point mass $M$ at the other extremum; and of a second
massless beam of length $r$, joined at one extremum with the
extremum of the first beam where the point mass is, having a point
mass $m$ at the other extremum.

\begin{figure}
  \includegraphics[width=150pt]{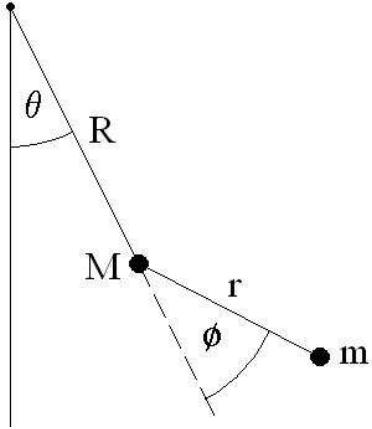} \\
  \caption{Notation for each of the double pendulums along the chain; see text for detail.}
  \label{dbpend}
\end{figure}

The pendulums can only rotate in the plane orthogonal to the axis.
The rotation angle of the first pendulum at site $i$ will be
denoted as $\th_i$; the angle of rotation of the second pendulum
(with respect to the direction of the first pendulum) will be
denoted as $\phi_i$. See fig.\ref{dbpend}.

The second pendulum is {\it not} free to swingle through a full
circle, but is instead constrained to stay in the range $|\phi^a_i
| \le \phi_0$. This constraint will also be modelled by adding a
constraining potential, i.e. a potential $V_c (\phi)$ which has
the effect of limiting {\it de facto} the excursion of the $\phi$
angles. The rest configuration will correspond to $\th_i = \phi_i
= 0$.

Let us denote, dropping for a moment the index $i$, as $(X,Y)$ and
$(x,y)$ the cartesian coordinates -- in the plane orthogonal to
the double helix axis $\xi$, with origin in the suspension point
-- of the point mass at the end of the first and of the second
pendulum respectively. The vertical direction will be along the
$x$ axis.

In the equilibrium position, given by $\th=\phi=0$, we have $(X,Y)
= (R,0)$ and $(x,y) = (R + r,0)$. In general, it will be
\beq\label{positions} \begin{array}{ll} X = R \cos \th \ , & x = R
\cos \th + r \cos (\th + \phi ) \ ; \\
Y = R \sin \th  \ , & y = R \sin \th + r \sin (\th + \phi ) \ .
\end{array} \eeq

The kinetical energy for the double pendulum at site $i$ will then
be $ T_i = (1/2) [ M  (({\dot X}_i)^2 + ({\dot Y}_i)^2 ) + m
(({\dot x}_i)^2 + ({\dot y}_i)^2) ]$; using (\ref{positions}) to
express $X_i,Y_i$ and $x_i,y_i$ in terms of the angles $\th_i$ and
$\phi_i$, we get
\beq\label{kin} \begin{array}{l} T_i \ = \
\frac{MR^2}{2} {\dot \th}_i^2  +\frac{m}{2} \left[ R^2 {\dot \th}_i^2 +
2 R r \cos \phi_i \( ({\dot \th}_i)^2 + {\dot \th}_i {\dot \phi}_i
\) +  r^2 ({\dot \th}_i + {\dot \phi}_i)^2 \right] \ .
\end{array} \eeq The total kinetic energy is of course $T = \sum_i
T_i$.

As for interactions, each pendulum will be coupled to nearest
neighbors via harmonic potentials, i.e. linear springs attached to
the $M$ and to the $m$ masses of the pendulums at sites $k$ and
$k+1$. Moreover, the mass $m$ at the end of the second pendulums
will experience a (site-independent) external potential.

\begin{figure}
  \includegraphics[width=150pt]{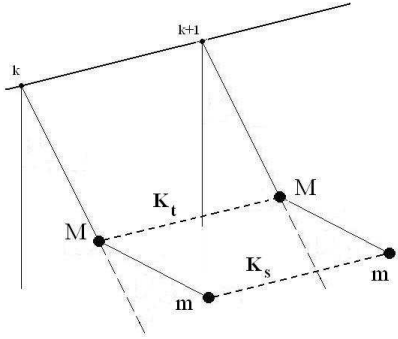}\\
  \caption{Pendulums at successive sites $k$ and $k+1$ interact via
  harmonic potentials coupling the masses at corresponding positions
  on the two pendulums; see text.}\label{dpendinter}
\end{figure}

Thus we have the following forces, and correspondingly potential
energy terms in the Lagrangian:

\medskip\noindent {\tt (1)} A coupling between successive
``first pendulums'', described by a potential \beq\label{pot}
U_t^{(i)} \ = \ \kappa_t \, [ 1 - \cos (\th_{i+1} - \th_i ) ] \ ;
\eeq this will be referred to as the ``torsional interaction''.
The total torsional potential is of course $V_t = \sum_{a,i}
U_t^{(a,i)}$.

\medskip\noindent {\tt (2)} A coupling between successive
``second pendulums'' on each chain, described by a potential $$
U_s^{(i)} = (1/2) \kappa_s [ (x_{i+1} - x_i)^2 + (y_{i+1}-y_i)^2 ]
\ ; $$ this will be referred to as ``stacking interaction''. In
terms of the $\th_i$ and $\phi_i$ angles, using again
(\ref{positions}) and with $\la = r/R$, we get \beq\label{stack}
\begin{array}{rl} U_s^{(i)}
 =&  ( \kappa_s R^2 / 2) \, \[ \( (\cos \th_i - \cos \th_{i+1}) +
 \la [ \cos (\phi_i + \th_i ) -
 \cos( \phi_{i+1} + \th_{i+1})] \)^2 \right.
\\ & \left. + \( (\sin \th_i - \sin \th_{i+1} ) + \la [ \sin(
\phi_i + \th_i)  - \sin( \phi_{i+1} + \th_{i+1} )] \)^2 \] \ ;
\end{array} \eeq the total stacking potential is of course $V_s =
\sum_{a,i} V_s^{(a,i)}$.

\medskip\noindent {\tt (3)} Interaction with the external
potential, \beq\label{pair} U_p^{(i)} \ = \ F ( \th_i , \phi_i ) \
. \eeq The total external potential will of course be $V_p =
\sum_{i} U_p^{(i)}$. In our case, the external potential will just
be the gravitational one, i.e. \beq\label{gravity} F (\th_i ,
\phi_i ) \ = \ g [M R (1-\cos \th_i)+ m[R+r- (R \cos \th_i + r
\cos (\phi_i + \th_i ))]] \ . \eeq

\medskip\noindent {\tt (4)} Interaction with the confining
potential, \beq\label{conf} U_c^{(i)} \ = \ h ( \phi_i ) \ , \eeq
with $h (x)$ a convex even function, essentially flat for $|x| \ll
 \phi_0$, and rising sharply at $|x| \approx \phi_0$.
 The total external potential will of course be $V_c =
\sum_{i} U_c^{(i)}$.

\bigskip

We have thus provided explicitly all terms appearing in the total
Lagrangian \beq\label{discrlag} L \ = \ T \, - \, \( U_t + U_s +
U_p \) \ . \eeq

The dynamics of the model will be governed by the corresponding
Euler-Lagrange equations, \begin{eqnarray}\label{dieulag} E_\th \
&:=& \ {\pa L \over \pa \th_i} \, - \, {\d \ \over \d t} \( {\pa L
\over
\pa {\dot \th}_i } \) \ = \ 0 \ , \nonumber \\
E_\phi \ &:=& \  {\pa L \over \pa \phi_i} \, - \, {\d \ \over \d
t} \( {\pa L \over \pa {\dot \phi}_i } \) \ = \ 0 \ .
\end{eqnarray}  We will not write these explicitly; they are
rather involved, due mainly to the kinetic term.\footnote{It would
be possible to obtain a simpler form for most terms by using
different coordinates, i.e. passing to $(\theta , \psi )$ with
$\psi$ the angle made with the rest (vertical downward) direction
by the line from the suspension point of the first pendulum to the
mass of the second pendulum. This would however introduce a very
complex dependence of the confining potential $h(\phi ) = \^h
(\theta , \psi )$, which we prefer to avoid.}

\section{Continuum version}

If one is mainly interested in solutions varying little on the
intersite scale, i.e. such that at all times $|\th_{i+1} - \th_i|
$ and $|\phi_{i+1} - \phi_i|$ are small -- say of order $\^\eps
\ll 1$, we can pass to a continuum description.

In this, the arrays $\th_i (t)$ and $\phi_i (t)$ will be replaced
by fields $\Theta (x,t)$ and $\Phi (x,t)$ such that, say, $\Phi (k
\de , t) = \phi_k (t)$ and $\Theta (k \de , t) = \th_k (t)$.

Expanding in a Taylor series up to order two in $\de$ we have
\beq\label{expand} \begin{array}{l} \Phi (x \pm \de , t) = \Phi
(x,t) \pm \de \Phi_x (x,t) + (\de^2/2) \Phi_{xx} (x,t) \ , \\
\Theta (x \pm \de , t) = \Theta (x,t) \pm \de \Theta_x (x,t) +
(\de^2/2) \Theta_{xx} (x,t) \ . \end{array} \eeq

Inserting this into the Euler-Lagrange equations (\ref{dieulag}),
we get the field equations, see (\ref{contEL}) below, governing
our system in the continuum approximation. In order to write
these, it is convenient to set \beq \a (\Phi) \ := \ 1 + (R/r)
\cos \Phi \ , \ \ \b (\Phi ) \ := \ (1 + (R/r)^2 + 2 (R/r) \cos
\Phi) \; \eeq we will moreover set $$ K_s \ := \ \kappa_s \, \de^2
\ , \ \ K_t \ := \ \kappa_t \, \de^2 \ . $$ With these, we get the
field equations \beq\label{contEL}
 \begin{array}{l} \Theta_{xx} [K_t + K_s r^2 \b (\Phi)] +
  \Phi_{xx} [K_s r^2 \a (\Phi)] \ = \\
  \ = \ \Phi_{tt} [m r^2 \a(\Phi)] +
  \Theta_{tt} [M R^2 + m r^2 \b (\Phi)] + \\
  \ \ -
  r R [m \Phi_t (\Phi_t + 2 \Theta_t) - K_s \Phi_x (\Phi_x + 2 \Theta_x)] \sin \Phi + \\
  \ \ + g [R (m + M) \sin \Theta + m r \sin (\Phi + \Theta )] \ ; \\
  K_s \Phi_{xx} r^2 + \Theta_{xx} [K_s r^2 \a(\Phi )] \ = \
  m \Phi_{tt} r^2 +
  \Theta_{tt} [m r^2 \a (\Phi)] + \\ \ \ + h'(\Phi) +
  r R (m \Theta_t^2 - K_s \Theta_x^2) \sin \Phi + g m r \sin (\Phi +
  \Theta )
  \ .  \end{array} \eeq

We will from now on work in the continuum approximation, i.e. with
(\ref{contEL}).

These are also obtained by rewriting the Lagrangian $L$ in the
continuum approximation; the explicit expression of this (written
for future reference) is \beq\label{contLag}
\begin{array}{l} \L = (1/2) \[ M R^2 \Theta_t^2 - K_t \Theta_x^2 +
m \( r^2 (\Phi_t + \Theta_t)^2 +R^2
\Theta_t^2 \right. \right.  \\
\left. \left. + 2 r R \Theta_t (\Phi_t + \Theta_t) \cos \Phi \) \right. \\
 \left. + 2 g [ (m + M) R \cos \Theta + m r \cos (\Phi + \Theta)
] - 2 h[\Phi] \right. \\ \left. - K_s [ \(R \Theta_x \cos \Theta +
r (\Phi_x + \Theta_x) \cos (\Phi + \Theta) \)^2 \right.
\\  \left. +
 \( R \Theta_x \sin \Theta + r (\Phi_x + \Theta_x) \sin(\Phi + \Theta) \)^2 ] \] \ .
  \end{array} \eeq

Then (\ref{contEL}) are also obtained as the associate
Euler-Lagrange equations
\begin{eqnarray}\label{eulag} \mathcal{E}_\Theta & := & {\pa \L \over \pa
\Theta} \, - \, {\d \ \over \d t} \( {\pa \L \over \pa \Theta_t }
\) \, - \, {\d \ \over \d x} \( {\pa \L \over
\pa \Theta_x } \) \ = \ 0 \ , \nonumber \\
\mathcal{E}_\Phi & := & {\pa \L \over \pa \Phi} \, - \, {\d \
\over \d t} \( {\pa \L \over \pa \Phi_t } \) \, - \, {\d \ \over
\d x} \( {\pa \L \over \pa \Phi_x } \) \ = \ 0 \ . \end{eqnarray}

The field equations (\ref{contEL}) should be supplemented by a
specification of the function space the solution are required to
belong to; these can be given as boundary conditions.

The physically natural condition is that of {\it finite energy} at
any time $t$; that is, considering the energy density \beq H (x,t)
\ = \ H[\Theta (x,t),\Phi(x,t)] \ = \ T + (U_t + U_s + U_p + U_c )
\ , \eeq and fixing the additive arbitrary constant so that the
minimum of the potential energy is at zero, the requirement that
\beq\label{finiteenergy} \int_{-\infty}^{+\infty} H (x,t) \, \d x
\ < \ \infty. \eeq

For this to be satisfied, in view of the form of $T$, we must
require that \beq \lim_{x \to \pm \infty} \Phi_x \ = \ \lim_{x \to
\pm \infty} \Theta_x \ = \ 0 \ , \eeq i.e. that $\Phi$ and
$\Theta$ are asymptotically constant in $x$.

Moreover, the fields $\Phi$ and $\Theta$ should go at minima of
the potential energy for $x \to \pm \infty$; in view of the form
of the potential energy, we must actually require that \beq
\lim_{x \to \pm \infty} \Phi \ = \ 0 \ ; \ \lim_{x \to \pm \infty}
\Theta \ = \ 2 \pi n_\pm \ . \eeq (Note that the difference
between topological and non topological degrees of freedom shows
up here.)

We can always change origin of the $\Theta$ angles, so that \beq
\lim_{x \to - \infty}  \Theta = 0 \ , \ \lim_{x \to + \infty}
\Theta = 2 \pi N \ . \eeq

Thus, finite energy solutions possess a topological index, the
integer $N$, identified by the asymptotic behavior at $x \to \pm
\infty$. One gets easily convinced that these boundary conditions
(and the finite energy condition) are preserved under the time
evolution described by (\ref{contEL}).

\section{Travelling wave solutions}

We are specially interested in travelling wave (TW) solutions,
i.e. solutions such that \beq\label{tw} \Theta (x,t) = \vth (x -
vt) \equiv \vth (z) , \ \Phi (x,t) = \vphi (x-vt) \equiv \vphi (z)
\ . \eeq

It is immediate to see that the boundary conditions implied by the
requirement of finite energy become in this framework
\beq\label{bc} \begin{array}{ll}
\lim_{z \to \pm \infty} \vphi'
(z) = 0 \ ,& \lim_{z \to \pm \infty} \vth' (z) = 0 \ ; \\
\lim_{z \to \pm \infty} \vphi (z) = 0 \ ,& \lim_{z \to \pm \infty}
\vth (z) = 2 \pi n_\pm \ . \end{array} \eeq We will again choose
variables so that $\vth (- \infty ) = 0$, $\vth (+ \infty ) = 2
\pi N$.

Inserting the ansatz (\ref{tw}) into the equations (\ref{contEL}),
and writing $M = \rho m$ for ease of notation, we get the
equations for travelling wave solutions: \beq\label{TWEL}
\begin{array}{l}
\mu \vphi_{zz} (r^2 + r R \cos \vphi) +
  \vth_{zz} [K_t - K_s R^2 \rho + \mu (r^2 + R^2 + R^2 \rho +
  2 r R \cos \vphi)] + \\
  \ \ -
  \mu r R \vphi_z^2 \sin \vphi - 2 \mu r R \vphi_z \vth_z \sin \vphi
  - m g [R (1 + \rho ) \sin \vth + r \sin (\vphi + \vth)] \ = \ 0 \ ; \\
  { } \\
\mu r^2 \vphi_{zz} + \mu \vth_{zz} r(r+ R \cos \vphi) -
h'(\vphi) +
  \mu r R \vth_z^2 \sin \vphi - m g r \sin (\vphi + \vth ) \ = \ 0 \ . \end{array}
  \eeq

Here we have simplified the writing by defining the parameter
\beq\label{mu} \mu \ := \ K_s - m v^2 \ ; \eeq this will have a
relevant role in the following.


In the following it will be of interest to consider the
Lagrangian $\L_{tw}$ producing the travelling wave equations
(\ref{TWEL}) as associated Euler-Lagrange equations; this is also
obtained by restricting the Lagrangian (\ref{contLag}) to the
space of functions satisfying (\ref{tw}). It results
\beq\label{lagtw}
\begin{array}{ll} \L_{tw} \ =& (1/2)
\[ m r^2 \b (\vphi ) v^2 + M
R^2 v^2 - K_t \right. \\
& \left. \ \ \ - K_s r^2 \b ( \vphi ) \] \,
\vth_z^2 \ + \ (1/2) \[ r^2 (m v^2 - K_s) \] \, \vphi_z^2 \\
 & \ + \[ (m v^2 - K_s) r^2 \a (\vphi) \] \, \vth_z \vphi_z \\
 & \ + \[ \( (m + M) g R \cos \vth + m g r \cos (\vphi + \vth)
 \) + h (\vphi) \] \ . \end{array} \eeq

\section{Series expansion for TW solutions}
\def\MTOT{{\^M}}
\def\MU{{\^\mu}}

A way to obtain the simple pendulums chain (and the
sine-Gordon equation in the continuum limit) from our model
is to let the length of second pendulums go to zero. Note that in
this case our set of parameters becomes redundant, as the masses
$m$ and $M$ coincide in space, so that only the total mass $\^M =
m+M$ is relevant; and similarly for the coupling constants $K_s$
and $K_t$, with total coupling strength $\^K = K_s + K_t$.

If the double pendulums chain is seen as a (singular) perturbation
of the simple pendulums one, one is naturally led to look for
travelling wave solutions as perturbations of the standard
sine-Gordon solitons.

We will thus now look for solutions to the (\ref{TWEL}) equations
in the form of a series expansion in a small parameter $\eps$,
\beq\label{seriestp}
\begin{array}{l}
\vth \ = \ \vth_0 + \eps \vth_1 + \eps^2 \vth_2 + .... \ , \\
\vphi \ = \ \vphi_0 + \eps \vphi_1 + \eps^2 \vphi_2 + .... \ .
\end{array} \eeq We will correspondingly also expand in the same
parameter the geometrical parameters and the masses appearing in
our model, and also allow for modification of the speed by
expanding it as well: \beq\label{seriespar}
\begin{array}{l} r \ = \ r_0 + \eps r_1 + \eps^2 r_2 + ... \ , \\
R
\ = \ A - r_0 - \eps r_1 - \eps^2 r_2 - ... \ ; \\
m \ = \ m_0 + \eps m_1 + \eps^2 m_2 + ... \ , \\
M \ = \ \MTOT - m_0 - \eps m_1 - \eps^2 m_2 - ... \ ;
\\
v \ = \ v_0 + \eps v_1 + \eps^2 v_2 + ... \ ; \end{array} \eeq
note that $A$ and $\MTOT$ represent the total length and total
mass of each double pendulum in the chain, which are kept constant
as $\eps$ is varied. In other words, we are now rescaling the
model with $\eps$.

We would like this is done so that for $\eps \to 0$ the double
pendulums chain reduces to a simple pendulums chain, with same
total length $A$ and total mass $\MTOT$; this requires $r_0 = 0$.
The choice of $m_0$ is at this point inessential (in the limit we
have a single pendulum with mass $\MTOT = M + m$), but for the
sake of simplicity we will choose $ m_0 = 0$.

It should also be noted that once the double pendulum reduces to a
simple one, the presence of two coupling constants $K_t$ and $K_s$
is actually redundant, as they intervene (in the limit, i.e. for
$\eps \to 0$) exactly in the same interaction: only $K := K_s +
K_t$ is relevant. Thus we will also write \beq\label{seriescoup}
\begin{array}{l}
K_t \ = \ \eps k_1 + \eps^2 k_2 + ... \ , \\
K_s \ = \ \^K \, - \eps k_1 - \eps^2 k_2 - ... \ . \end{array}
\eeq

Finally, we note that for $\eps \to 0$ and hence for the single
pendulums chain, the angle $\vphi$ makes no sense; in order to
avoid any paradoxical behavior, we would require that it gets
frozen to zero for $\eps \ll 1$, i.e. \beq\label{phizero} \vphi_0
\ = \ 0 \ . \eeq

Summarizing our discussion, the series expansion we adopt are as
follows: \beq\label{seriesall}
\begin{array}{lll}
\vth  & = \  \vth_0 + \eps (\de \th)  & = \  \vth_0 + \eps \vth_1 + \eps^2 \vth_2 + .... \ , \\
\vphi  & = \  \eps (\de \vphi)  & = \  \eps \vphi_1 + \eps^2 \vphi_2 + .... \ ; \\
r  & = \  \eps (\de r)  & = \  \eps r_1 + \eps^2 r_2 + ... \ , \\
R  & = \  A - \eps (\de r)  & = \  A  - \eps r_1 - \eps^2 r_2 - ... \ ; \\
m  & = \  \eps (\de m)  & = \  \eps m_1 + \eps^2 m_2 + ... \ , \\
M  & = \  \MTOT - \eps (\de m)  & = \  \MTOT - m_0 - \eps m_1 - \eps^2 m_2 - ... \ ; \\
v  & = \  v_0 + \eps (\de v)  & = \  v_0 + \eps v_1 + \eps^2 v_2 + ... \ ; \\
K_t  & = \  \eps (\de \^K)  & = \  \eps k_1 + \eps^2 k_2 + ... \ , \\
K_s  & = \  \^K - \eps (\de \^K)  & = \  \^K \, - \eps k_1 -
\eps^2 k_2 - ... \ . \end{array} \eeq

We will then insert the series expansions (\ref{seriesall}) in the
Euler-Lagrange equations for TW solutions (\ref{TWEL}). It will be
convenient to define \beq \^\mu \ := \ ( \^M \, v_0^2 \ - \ \^K )
\ . \eeq

\subsection{Terms of order zero}

At order zero, the second equation in (\ref{TWEL}) is identically
satisfied (it just reduces to $h'(0) = 0$, which always holds for
$h$ even), while the first reads \beq\label{th0eq} {\vth_0}'' \ =
\ - {\^M g \over A \^\mu } \ \sin (\vth_0 ) \ := \ - \, \kappa \
\sin (\vth_0 ) \ . \eeq

This reproduces -- as obvious by construction -- the sine-Gordon
equations for the single pendulums chain. Its solution is
\beq\label{theta0} \vth_0 (z) \ = \ 4 \ \arctan
\[ \exp ( \kappa \, z ) \] \ . \eeq
Derivatives of this function are readily computed, and we have
\beq \vth_0' \ = \ 2 {\kappa \over \cosh (\kappa z) } \ , \ \
{\vth_0}'' \ = \ - \, 2 \ {\kappa^2 \ \sinh (\kappa^2 z) \over
\cosh^2 (\kappa^2 z)} \ . \eeq These will appear in the higher
order term of the expansion of (\ref{TWEL}); in these we also find
terms like $\sin (\vth_0 )$ and $\cos (\vth_0 )$, which in view of
(\ref{theta0}) -- and with some simple algebra -- are given by
\beq \sin (\vth_0 ) \ = \ - 2 \, {\sinh (\kappa x) \over
\cosh^2(\kappa x)} \ , \ \ \cos (\vth_0 ) \ = \ {\sinh^2 (\kappa
x) - 1 \over \cosh^2 (\kappa x) } \ . \eeq

\subsection{Terms of order one}
\def\sech{{\rm sech}}

At order $\eps$, and using the equation at order $\eps^0$, the
first of (\ref{TWEL}) reads \beq \begin{array}{l} {\vth_1}'' \ = \
\ [A^2 \^\mu ]^{-1} \ \left[ \( (A^2 -1 ) k_1 + 2 A \^M
v_0 (A v_1 - r_1 v_0 ) \) {\vth_0}'' + \right. \\
\left. \ \ + (A \^M g \cos \vth_0 ) \vth_1 - \^M g r_1 \sin \vth_0
\right] \ . \end{array} \eeq
Using the solution for $\vth_0$ and
its consequences recalled above, this simplifies to
\beq\label{eq:ord1a}
\begin{array}{ll}
{\vth_1}'' \ =& - \, \( \^M^2 g^2 \sech^2 (\kappa z) \, / \, ( 2
A^5 \^K^2 )  \) \ \times \\ & \times  \[ A^3 \^K \( \cosh (2
\kappa z) - 3 \) \, \vth_1 \ + \ 4 \, B \,  \, \sinh (\kappa z)
\] \ , \end{array} \eeq
where we have written \beq B \ := \ \( A^2 \kappa r_1 - \^M g ((1
- A^2 ) k_1 + 2 A \^M v_0 (r_1 v_0 - A v_1 ))\) \ . \eeq

As for the order $\eps$ terms in the second of (\ref{TWEL}), these
yield \beq\label{eq:ord1b} \vphi_1 \ = \ {\^M g \^K r_1 \over
\^\mu \, h''(0)} \ \sinh (\vth_0 ) \ . \eeq

It should be stressed that these two equations, (\ref{eq:ord1a})
and (\ref{eq:ord1b}), are qualitatively different: in facts, while
the first is a (second order) differential equation for $\vth_1$,
which depends on $\vth_0$, the second is an algebraic identity
which directly determines $\vphi_1$.

\subsection{Higher order terms}

Proceeding with the analysis at higher and higher orders, we would
of course obtain more and more involved explicit equations.
However, the main feature displayed at order one will be present
at higher orders $k > 1$ as well.

That is $\vth_k$ will be determined by a differential
equation of the form
$$ {\vth_k}'' \ = \ F_k [\vth_0,...,\vth_{k-1}; \vphi_1,...,\vphi_{k-1} ] \,
\vth_k \ + \  G_k [\vth_0,...,\vth_{k-1}; \vphi_1,...,\vphi_{k-1}
] \ $$ (omitting for the sake of brevity the dependence on
parameters).

This can be seen as the equation of motion for a particle of unit
mass, whose position is described by $\vth_k (t)$, in the
time-dependent potential \beq V_k (\vth_k , t) \ := \ - (1/2)
\widetilde{F}_k (t) (\vth_k)^2 \ - \ \widetilde{G}_k (t) (\vth_k)
\ ,  \eeq where we have used the fact $\vth_j (t)$, $\vphi_j (t)$
with $j < k$ can be determined by solving equations at lower
orders so to write\footnote{This obviously requires to solve the
equations order by order, as always in perturbative approaches.}
 \beq \widetilde{F}_k (t) \ := \ F_k [\vth_0
(t),...,\vth_{k-1} (t); \vphi_1 (t),...,\vphi_{k-1} (t)] \ . \eeq

As for the term $\vphi_k$, this will {\it not } be determined by a
differential equation, but rather by an algebraic relation, i.e.
\beq\label{eq:slaving} \vphi_k \ = \ H_k [\vth_0,...,\vth_{k-1};
\vphi_1,...,\vphi_{k-1} ]  \ . \eeq

E.g., at order $\eps^2$, and using the results from the analysis
of terms of order $\eps^0$ and $\eps^1$, we get for this algebraic
relation\footnote{The only interest of such an involved formula is
to show that the algebraic relation can be explicitly determined
at each order, by standard recursive computations.} \beq
\begin{array}{l} \vphi_2 = [g /(2 A^2 \^\mu^2 h'' (0))]
\times \\ \ \times
\[ A^2 K \^M r_1 \sech^2 (\kappa z) (\cos ( 2 \kappa z) -3) (h''
(0))^2 \vth_1 \right. \\ \ \left. + \  4 K \^M r_1 [k_1 - A^2 k_1
+ A ( \^\mu r_1 + 2 \^M v_0 (r_1 v_0 - A
v_1))] \times \right. \\
 \ \left. \times \sech (\kappa z) \tanh (\kappa z) (h''(0))^2 \right. \\
\ \left. + \ A^2 \sin (\vth_0) [- 2 \^\mu h''(0) (- A K^2 \^M
r_1^2 (\vth_1')^2 \right. \\ \ \left. + \ (k_1 \^M r_1 - K \^M r_2
+ m_1 r_1 (\^\mu + \^M v_0^2 )) h''(0) ) \right. \\ \ \left. - \ g
K^2 \^M^2 r_1^2 \sin (\vth_0 ) h'''(0) ]
\] \ . \end{array} \eeq

It should be noted that the algebraic -- rather than differential
-- character of the relation (\ref{eq:slaving}) is due, in
algebraic terms, to the choice $r_0 = 0$ (which also means we are
considering a {\it singular perturbation} \cite{Ver}). This makes
that the terms with derivatives of $\vphi_k$ will {\it not }
appear in the $O(\eps^k )$ terms of the expansion, while $\vphi_k$
itself does; hence we have an algebraic relation and not a
differential equation.

In physical terms, the choice $r_0 = 0$ corresponds indeed to
considering the double pendulums chain as a perturbation of the
single pendulums one, the perturbation parameter being related to
the length of second pendulums.

In facts, let us substitute according to the condensate form of
(\ref{seriesall}) into (\ref{lagtw}): we write the resulting
expansion as \beq\label{LTWpert} \L_{tw} \ = \ \L_0 \ + \ \eps \,
\L_1 \ + \ \eps^2 \, \L_2 \ + \ O (\eps^3) \ . \eeq We will {\it
not } set $\vphi_0 \equiv 0$. At first orders we have the
expressions reported in Appendix B

The complete explicit form of these is not so relevant; the
important fact is that -- as can be checked explicitly by the
formulas in Appendix B -- we have \beq \label{e1} \pa \L_k / \pa
\vphi_k' \ = \ 0 \ ; \eeq hence the corresponding Euler-Lagrange
equation (no sum on $k$ here and below) \beq\label{e2} {\pa \L_k
\over \pa \vphi_k} \ - \ {d \over d t} {\pa \L_k \over \pa
\vphi_k'} \ = \ 0 \eeq reduces to the identity \beq\label{ELident}
\A_k \ := \ \ {\pa \L_k \over \pa \vphi_k} \ = \ 0 \ . \eeq

These equations show that -- as anticipated in the Introduction --
to any order of the perturbative expansion the Lagrangian $\L_k$
depends on $\vphi_k$ but is independent of the momentum conjugated
to $\vphi_k$: that is, the $\vphi_k$ can be considered as an
auxiliary field, entering in the Lagrangian only algebraically
(and not differentially). We express this fact by saying that the
field $\vphi$ is a {\bf perturbatively auxiliary field}. We stress
that $\vphi$ is not an auxiliary field in the full Lagrangian,
i.e. its auxiliary character arises as a consequence of the
(singular) perturbative expansion we considered.

In facts, the above explicit expressions for $\L_0,\L_1,\L_2$
provide the expression for the first few identities $\A_k$, which
all reduce to $h' (\vphi_0) = 0$.

If we look at the Euler-Lagrange equations \beq {\mathcal E}_{k,j}
\ := \ {\pa \L_k \over \pa \vphi_{k-j}} \ - \ {d \over d t} {\pa
\L_k \over \pa \vphi_{k-j}'} \ = \ 0 \ , \eeq these reduce to the
algebraic identities seen above.

In particular, we have \begin{eqnarray*} {\mathcal E}_{1,0} &=& A
K r_1 (\vth_0'' \cos \vphi_0 + 2 (\vth_0')^2
\sin \vphi_0 ) + \vphi_1 h'' (\vphi_0 ) = 0 \ ; \\
{\mathcal E}_{2,1} &=& A K r_1 (\vth_0'' \cos \vphi_0 + 2
(\vth_0')^2
\sin \vphi_0 ) + \vphi_1 h'' (\vphi_0 ) = 0 \ ; \\
{\mathcal E}_{2,0} &=& [ - K r_1^2 \vth_0'' + A ( K ( 2 \vphi_1
r_1 (\vth_0')^2 + r_2 \vth_0'' + r_1 \vth_1'' ) \\ & & \ \ \ - r_1
(k_1 + m_1 v_0^2) \vth_0'' )] \cos \vphi_0 \\ & & - [2 K r_1^2
(\vth_0')^2 + A (2 k_1 r_1 (\vth_0')^2 - 2 K r_2 (\vth_0')^2 + K
r_1 \vphi_1 \vth_0'' - 4 K r_1 \vth_0' \vth_1' \\ & & \ \ \ + 2
m_1 r_1 v_0^2 (\vth_0')^2 )] \sin \vphi_0 \\ & &    + r_1 (K r_1
(2 \vphi_0'' + \vth_0'' ) - g m_1 \sin (\vphi_0 + \vth_0 ) ) +
(1/2) \vphi_1^2 h''' (\vphi_0 ) \\ & & + h''(\vphi_0 ) \, \vphi_2
\ = \ 0 \ . \end{eqnarray*}

\section{Discussion and conclusions}

We have considered a model corresponding to a chain of double
pendulums (with a constraining potential limiting excursions of
second pendulums), with first-neighbor harmonic interactions and
possibly subject to an external potential. This can, in an
appropriate limit, be considered as a perturbation of the usual
simple pendulums chain with harmonic interactions, possibly
subject to an external potential, leading to (discrete)
sine-Gordon solitons.

We dealt with these models in the continuum limit; in this case
the field equations we obtain describe two interacting fields
(possibly in  an external potential) a topological field $\Theta
(x,t)$ and a non-topological one $\Phi (x,t)$.

Looking for travelling wave solutions -- which include in
particular simple solitons -- with velocity $v$, we obtain a
reduction to two coupled second order ODEs for $\vth (z)$ and
$\vphi (z)$, where $z = (x - v t)$.

These equations posses a Hamiltonian structure and as such they
can be studied using conservation of energy; this reduces the
effective dynamics -- for given initial data -- to a three
dimensional manifold. However, the lack of a second constant of
motion prevents integrability and hence obtaining a general
solution for the dynamics.

We argued then that our model can be seen, as mentioned above, as
a perturbation of the standard simple pendulums model leading in
the continuum approximation to sine-Gordon equations. The
perturbation parameter $\eps$ should lead not only field
expansion, but also appropriate expansions for the geometrical and
dynamical parameters of the model. We would thus expect that --
for small $\eps$ -- the solitons of our model can be expressed as
perturbations of the familiar sine-Gordon solitons.

By performing explicitly the perturbative expansion and solving
the resulting equation at first order, we have shown that this is
the case. However, in this procedure we obtained a rather
unexpected feature: that is, the non-topological field is
determined by the topological one via an algebraic relation --
{\it not } a differential equation -- and thus has the role of a
{\it slaved field}, while the topological field takes the role of
{\it master field}.

This feature is due to the appearance of a {\it perturbatively
auxiliary field}; that is, when we expand perturbatively our model
around the $\vth_0$ soliton, the field $\vphi = \eps \vphi_1 +
\eps^2 \vphi_2 + ... $ is an auxiliary field, determined by
algebraic equations, up to any desired order -- but it is {\it not
} an auxiliary field for the full dynamics.

We have already noted (see appendix A for details) that the simple
pendulum system can be also obtained, acting on the confining
potential, as a dynamical limit  of the double pendulums chain.
Thus, the double pendulums chain can be reduced to a single
pendulums chain in two conceptually (and physically) independent
ways: the dynamical way discussed in Ref. \cite{CDGspeed} and the
singular limit discussed in this paper. Although leading to the
same simple physical system, the two different reductions have
peculiar features that  make them rather different. Therefore they
may be used   in different contexts to model the dynamics of
realistic systems such as molecular chains.

The most striking difference between the two simple pendulums
reductions is related to the breaking of boost (Lorentz) symmetry.
Owing to the presence of two different sound speeds (related to
the presence of two coupling constants $K_{t},K_{s}$), the
lagrangian (\ref{contLag}) is not invariant under Lorentz
transformation \cite{CDGspeed}. However,  performing the $\eps \to
0$ limit, at the zeroth order in the perturbation  theory we
recover Lorentz invariance and the speed of the sine-Gordon
soliton is not fixed. The recovering of the boost symmetry  is
essentially a consequence of the fact that in the limit $\eps \to
0$ only the coupling $K=K_{s}+K_{t}$ and the mass $\hat M= m+M$
(not the single couplings or single masses) are relevant. It
follows that in this limit we have just one (and not two) speed of
sound $\sqrt{K/\hat M}$. Obviously, the boost symmetry is broken
by higher orders in the perturbative $\eps$-expansion, but
perturbative solution can be still found for any value of the
speed of the travelling wave. Conversely, in the case of  the
dynamical single pendulum reduction the Lorentz symmetry remains
broken also after the reduction and the soliton speed is fixed
(see appendix A). Moreover, in this later  case it is very
difficult to find  perturbative solutions of the double pendulums
chain and one has to resort to numerical calculations
\cite{CGDlong}.

In conclusion, the dynamical reduction to a single pendulums chain
provide us with a nice mechanism to fix the speed of sine-Gordon
solitons,  can be applied in presence of a strong confining
potential  but is not suitable for a perturbative  evaluation of
the solutions of the double pendulums model. On the other hand,
the singular $\eps$-expansion discussed in this paper provide us
with a nice perturbative framework for evaluating the general
solutions of the model, can be applied in presence of weak
confining potentials but does not allow for a soliton speed fixing
mechanism.

\section*{Appendix A. \\ The dynamical simple pendulums limit}

Now we note that if we force $\Phi (x,t) = 0$ -- and hence $\vphi
(z) = 0$ -- in our model, then we are actually considering a chain
of simple pendulums; in the continuum limit, this leads to a
sine-Gordon equation.

The constraint $\Phi (x,t)=0$ can be accommodated, in our setting,
by acting on the confining potential $U_c$, i.e. on $h$: this
should be made stronger and stronger -- hence we speak of a {\it
dynamical limit}, as opposed to the geometrical one considered
above -- and the maximum angle $\phi_0$ smaller and smaller.

In the limit\footnote{We stress that this is a {\it singular
limit}, as the limiting model ($\phi_0 = 0$) has a smaller number
of degrees of freedom than any other case ($\phi_0 \not= 0$).}
$\phi_0 \to 0^+$ and $h'' (0) \to + \infty$, we expect to recover
the solitons of the sine-Gordon equation. Note that in this limit
the coupling constants $K_s$ and $K_t$ actually refer to the same
interaction; we will thus take $K_t = 0$, also in order to make
comparison with the sine-Gordon case more immediate.

If we force $\vphi (z) \equiv 0$, and use $h' (0) = 0$, $K_t = 0$,
then the equations (\ref{TWEL}) reduce to $$
\begin{array}{l}
[\mu (2 r R + r^2 + R^2 (1 + \rho)) - K_s R^2 \rho] \vth_{zz} +
  g m (r + R + R \rho) \sin \vth \ = \ 0 \ ; \\
\mu r (r + R) \vth_{zz} + g m r \sin \vth \ = \ 0 \ .
\end{array} \eqno(A.1) $$

Each of these two equations provide a different determination for
$\vth_{zz}$; the compatibility condition for them is obtained
requiring the two determinations coincide. When $r \not= 0$ (if
$r=0$ the second equation is identically satisfied) the
compatibility condition reads $$  \mu \ = \ - K_s R / r \ . \eqno
(A.2) $$ This provides a unique determination for $\mu$ and hence,
in view of (\ref{mu}), for $v^2$; more precisely, this yields $$ v
\ = \ \pm \ \sqrt{{K_s (r + R) \over m r}} \ . \eqno(A.3) $$

Thus, {\it the compatibility condition selects uniquely the speed
of the travelling wave} obtained in the dynamical simple pendulums
limit.

This phenomenon is not peculiar to our model, but common to a
large class of multi-fields soliton models, as discussed elsewhere
\cite{CDGspeed}.

\section*{Appendix B. \\
Expansion of the travelling wave Lagrangian}

In this Appendix we provide explicit formulas for the first terms
in the perturbative expansion (\ref{LTWpert}) of the travelling
wave Lagrangian (\ref{lagtw}). These are:
$$\begin{array}{l} \L_0 = (A/2) \[ A (\^M v_0^2 - K) (\vth_0')^2 -
2 g \^M \cos \vth_0 - 2 h(0) \] \ ; \\ {} \\
\L_1 = \[ (1/2) (A^2 - 1) k_1 + A r_1 (K - \^M v_0^2) + A^2 \^M
v_0 v_1 - A K r_1 \cos \vphi_0 \] (\vth_0')^2 \\ \ - \ A^2 (K -
\^M v_0^2 ) (\vth_0' \vth_1' ) - \ A K r_1 \cos \vphi_0 (\vth_0'
\vphi_0') \\ \ - \ g \^M (a \vth_1 \sin \vth_0 - r_1 \cos \vth_0)
\vth_0' \ - \ [h' (\vphi_0 ) ] \vphi_1 \ ; \\ {} \\
\L_2 = (A^2/2) (\^M v_0^2 - K) (\vth_1')^2 - (1/2) K r_1^2
(\vphi_0')^2  - A K r_1 \cos \vphi_0 (\vphi_0' \vth_1') \\ \
 + (1/2) [ (A^2 -1) k_2 - 2 r_1 (A k_1 - K r_1) + 2 A K r_2
+ \^M r_1^2 v_0^2 - 2 A \^M r_2 v_0^2 \\ \ - 4 A \^M r_1 v_0 v_1 +
A^2 \^M v_1^2 + 2 A^2 \^M v_0 v_2 \\ \ \ + 2 (A k_1 r_1 + K r_1^2
- 2 A K r_2 + 2 A m_1 r_1 v_0^2) \cos \vphi_0 + 2 A K r_1 \vphi_1
\sin \vphi_0 ] (\vth_0')^2 \\ \ + \^M g r_2 \cos \vth_0 + (1/2) A
g \^M \vth_1^2 \cos \vth_0 - g m_1 r_1 \cos (\vphi_0 + \vth_0 )
\\ \ - g \^M (r_1 \vth_1 - A \vth_2) \sin \vth_0 -
\vphi_2 h'(\vphi_0) - (1/2) \vphi_1^2 h''(\vphi_0) \\
\ \ + [ A^2 (\^M v_0^2 - K) \vth_2' + ((A^2 - 1 ) k_1 + 2 A K r_1
- 2 A \^M r_1 v_0^2 + 2 A^2 \^M v_0 v_1 \\ \ \ \ - 2 A K r_1 \cos
\vphi_0 ) \vth_1' \ + \ ((A K_1 r_1 + K r_1^2 - A K r_2 + A m_1
r_1 v_0^2 ) \cos \vphi_0 \\ \ \ + A K r_1 \vphi_1 \sin \vphi_0 - K
r_1^2 ) ] \vth_0' \ .
\end{array} $$

These expressions allow to check explicitly (at first orders) that
(\ref{e1}) holds, and hence (\ref{e2}) reduce to (\ref{ELident}).


\end{document}